# Stumbling over planetary building blocks: AU Microscopii as an example of the challenge of retrieving debris-disk dust properties

Jessica A. Arnold[1], Alycia J. Weinberger[2], Gorden Videen[1,3,4], Evgenij S. Zubko[4]

[1]Army Research Laboratory 2800 Powder Mill Rd, Adelphi, MD 20783 (*jessy.arnold@gmail.com*), [2]Department of Terrestrial Magnetism, Carnegie Institution for Science, 5421 Broad Branch Rd. Washington, DC 20015, USA, [2]Space Science Institute, 4750 Walnut Street, Boulder Suite 205, CO 80301, USA, [3]Kyung Hee University, 1732, Deogyeong-daero, Giheung-gu, Yongin-si, Gyeonggi-do 17104, Republic of Korea

## Abstract

We explore whether assumptions about dust grain shape affect resulting estimates of the composition and grain size distribution of the AU Microscopii (AU Mic) debris disk from scattered light data collected by Lomax et al. (2018). The near edge-on orientation of the AU Mic debris disk makes it ideal for studying the effect of the scattering phase function (SPF) on the measured flux ratios as a function of wavelength and projected distance. Previous efforts to model the AU Mic debris disk have invoked a variety of dust grain compositions and explored the effect of porosity, but did not undertake a systematic effort to explore a full range of size distributions and compositions to understand possible degeneracies in fitting the data. The degree to which modelling dust grains with more realistic shapes compounds these degeneracies has also not previously been explored. We find differences in the grain properties retrieved depending on the grain shape model used. We also present here our calculations of porous grains of size parameters $x$ = 0.1 to 48 and complex refractive indices ($m = n+ik$) ranging from $n$ = 1.1 to 2.43 and $k$ = 0 to 1.0, covering multiple compositions at visible and near infrared wavelengths such as ice, silicates, amorphous carbon, and tholins.

## 1. Introduction

*Background*

AU Mic is a 23(+/-3) Myr old (Metchev, Eisner & Hillenbrand 2013; Mamajek & Bell 2014) M-type star located 9.72 pc from Earth. AU Mic resides in the Beta Pictoris moving group, whose age has been variously estimated at 10-40 Myr (Table 1 in Mamajek and Bell 2014), and most recently placed at 18.5 Myr in a re-analysis by Miret-Roig et al. (2020). The star is surrounded by a nearly edge-on (i = 89.5°) debris disk comprised of a cold, outer dust belt at 8.8-43 AU (MacGregor et al. 2013).

In optical scattered-light images, the outer disk halo extends to 210 AU (Strubbe and Chiang 2006). AU Mic is unusual because detections of disks around M stars are sparse relative to those around A-type or solar-type stars (Matthew et al 2014). The AU Mic debris disk was first spatially resolved in 2004 (Kalas et al. 2004; Liu et al. 2004) and has since been observed with a number of instruments and techniques, including Hubble Space Telescope (HST) and Keck II optical to near infrared scattered light (Krist 2005; Metchev et al. 2005; Fitzgerald et al. 2006; Lomax et al 2018), HST optical polarized scattered light (Graham et al. 2007), as well as Herschel Space Observatory (HSO) far-IR (Matthews et al. 2015), James Clerk Maxwell Telescope (JCMT) sub-mm (Matthews et al. 2015), and the Atacama Large Millimeter Array (ALMA) mm emission (Wilner et al. 2012; MacGregor et al. 2013; Daley et al. 2019). Clump-like features of variable brightness have been observed moving through the disk on timescales of months to years (Boccaletti et al. 2015; Chiang and Fung 2017). A close-in (0.07 AU), Neptune-sized (0.4 Jupiter radii) planet within the disk, AU Mic b, has recently been confirmed (Plavchan et al. 2020), a possibility that had been raised previously (e.g., Ertel et al. 2012b; Nesvold & Kuchner 2015). The potential dynamical relationships between the planet, the inner disk, and the moving outer disk structures have yet to be explored. Finding the properties of the outer dust belt will help us to understand the processes responsible for dust transport within the disk and characterize the environment in which AU Mic b formed.

Due to the cold average temperatures (Matthews et al. 2015) of the dust belt (39 K) and halo (50 K), it is not possible to study the 8-70-micron silicate features, since the far-infrared peak of the SED is near 100 microns. Therefore, the only possible way to extract compositional information is from the brightness and wavelength-dependent albedo (or color) of the disk in scattered light. The edge-on orientation of the AU Mic debris disk means that light scattered at different projected separations from the star will originate from different sets of scattering angles. This allows information about the dust to be extracted from the SPF, which describes how much light is scattered by the dust as a function of phase angle. However, both the SPF and resulting color of the disk at various projected separations are functions of dust grain size, shape, and porosity, in addition to composition. This means that compositional information can only be extracted if this degeneracy can be broken.

In scattered light, the AU Mic debris disk shows an overall blue color in the B through J bands. Several works have reported a change in color with projected separation from the central star (Krist et al. 2005; Fitzgerald et al. 2007; Lomax et al. 2018). However, the observed radial color trend is not consistent across wavelengths, and while several works report a bluing trend at projected separations of 30 AU, the color trend inside of that radial is not consistent between observations at similar wavelengths, likely due to differing spatial resolutions of the observations.

Here, we explore a range of compositions and dust grain-size distributions that simultaneously fit the scattered-light data at different projected separations over multiple wavelengths using an Markov chain Monte Carlo (MCMC) model. We use scattering properties computed for three different models: compact spheres, porous spheres and a more realistic agglomerated debris model.

*Observations of the optical to near-IR color and polarization of AU Mic*

Several works have investigated trends in band ratios or multi-wavelength observations of AU Mic at varying projected distances. These trends are outlined below and summarized in Table 1.

Krist et al (2005) presented multicolor HST Advanced Camera for Surveys (ACS) coronographic images of the AU Mic debris disk. They found an overall grey color when looking at the F435W/F606W ratio at a value of 1.02, but an overall blue color when looking at the F435W/F814W ratio at a value of 1.16. They also observed the SE side of the disk to be slightly bluer than the NW side. The F435W/F814W ratio appeared to be greater at 60 AU than 30 AU and also greater at <30 AU than at 30 AU. Krist et al. warned however, that the results at <30 AU may have been biased by the PSF subtraction residual of the 814 W band. They did not observe a similar radial trend in the F435W/F606W ratio.

Metchev et al. (2005) collected H band observations with Keck II AO and compared these to the R band measurements from Kalas et al. (2004). The F606W/H band color was neutral at 17–20 AU and blue 50–60 AU, agreeing with the radial trend observed by Krist et al. (2005). Metchev et al. (2005) also modeled the minimum and maximum grain size using the disk mass derived from the sub-mm JCMT/SCUBA SED and fixing the grain size distribution at $n(a) \sim a^{-3.5}$. They also used a set composition of 65% silicate plus graphite. From these fixed parameters and the scattered light data, they retrieved a minimum size of 1.6 um at 20 AU, 0.5 um at 50-60 AU and a maximum grain size of 300 μm. A break in the surface brightness profile was observed in the outer disk (>33 AU).

Bluing towards the outer disk (30-60 AU) was also observed in Keck II J and H band observations by Fitzgerald et al. (2007), but this trend did not extend to the K band. Similar to Metchev (2005), a break in the surface brightness profile at 30-35 AU was observed. They modeled the grain size distribution of the outer disk using a porous grain model with 80% porosity made up of a mixture of silicates, carbonaceous materials, and ice. They compared their porous grain model to a compact silicate only model. Fixing the power-law exponent for the inner disk at $q = 3.5$, where $n(a) \sim a^{-q}$, they found the outer disk has a power-law exponent of $q = 4.1$. They were able to reproduce the scattered light and far-IR SED with an inner zone (35–40 AU) of few-mm-sized grains and an outer zone with smaller 50 nm–3.0 μm grains.

A two-zone power law consisting of porous, icy grains was also favored by Graham et al. (2007) to match their observations of HST-ACS polarimetry at F606W using the POLV polarizing filter. Graham et al. (2007) reported an increase in the degree of linear polarization from 0.05 to 0.35 at 20-50 AU, followed by a leveling off at 50-80 AU. They tried several models including two different Henyey-Greenstein functions, compact spheres composed of dirty ice, compact spheres composed of silicates, zodiacal dust and ISM phases functions, but the best fit was given by >90% porosity icy aggregates approximated using the Maxwell-Garnett mixing rule and Lorenz-Mie scattering. They note that while small spheres can replicate the polarization data, they cannot also simultaneously fit the surface brightness profile as they are not forward scattering enough. The lower values of the degree of linear polarization in the inner disk was interpreted by Graham et al. (2007) as resulting from a depletion of small grains in that region.

Higher spatial and spectral resolution HST Space Telescope Imaging Spectrograph (STIS) observations collected at several projected separations ranging from 10 to 45 AU were presented by Lomax et al. (2018). These data showed the interior of the disk to be bluer, with the spectral slope becoming increasingly grey out to 45 AU. While the observations likely did not include enough of the halo to compare with previous observations that noted a bluing trend outward of 30 AU, there is an apparent discrepancy with observations suggesting that the disk is spectrally neutral in this wavelength regime interior to 30 AU. However, a blue slope interior to 30 AU does agree with the trend reported by Krist (2005) and could be due to the fact that the other studies binned large areas in projected separation.

While some of the studies mentioned above included attempts to model the observed changes in disk albedo and polarization with radial distance, they did so using a limited number of compositions with fixed proportions of silicates, ice, and carbon. This means the range of possible compositions has not until this point been explored, making it unclear how unique these solutions are.

**Table 1:** Summary of color and linear polarization trends observed in the AU Mic debris disk for different projected separations.

| Study | Instruments | Measurements | Projected Separations | Trends |
|---|---|---|---|---|
| Krist et al (2005) | HST-ACS | F435W, F606W, F814W | Whole disk | Grey F435W/F606W; Blue F435W/F814W ratio |
| | | | <30 AU | Bluer F435W/F814W |
| | | | 30-60 AU | Less blue F435W/F814W |
| | | | >60 AU | Bluer F435W/F814W |
| Metchev et al. (2005) | Keck II AO | H band (c.f. F606W from Kalas et al.) | 17–20 AU | Grey F606W/H ratio |
| | | | 50–60 AU | Blue F606W/H ratio |
| | | | 33 AU | Break in surface brightness profile |
| Fitzgerald et al. (2007) | Keck II | J, H, K bands | 30–35 AU | Break in surface brightness profile |
| | | | 30–60 AU | Outward bluing trend observed in J, H, but not K |
| Graham et al. (2007) | HST-ACS | F606W with POLV filter | 20–50 AU | Increasing linear polarization |
| | | | 50–80 AU | Leveling-off of linear polarization |
| Lomax et al. (2018) | HST-STIS | Multi-spectral 0.52-1.02 μm | 10–45 AU | Blue interior at 12-17 AU, gradual greying out to 45 AU |

*Observations and models of grain dynamics*

One consistent interpretation of the overall blue color of the disk is that the dust is dominated by small (<1 μm) grains. This is consistent with the fact that the radiation pressure force is too weak around an M-type star to effectively remove small grains from the system (Arnold 2019). Augereau and Beust (2006) found that even accounting for X-ray and UV flares, radiation pressure is still inefficient. However, they found that corpuscular pressure from the stellar wind (Plavchan et al. 2005) could diffuse small grains (<1 um) to the outer disk. Interior to 30 AU the drag component of the stellar wind becomes important and may explain the relative lack of dust in that region.

Strubbe and Chiang (2006) modelled the dust dynamics of grains generated by collisions of objects in the birth ring of planetesimals at 43 AU. The Poynting-Robsertson drag timescale is a function of stellar-centric distance, but is long, and that stellar wind drag is likely the limiting factor for dust grain size. They posit that the dust generated within the parent belt smaller than the size dominated by collisional removal, but larger than the stellar wind and radiation-pressure removal size, gets dragged inward due to stellar wind and grains just larger than the stellar wind/radiation pressure threshold are sent into highly elliptical orbits beyond the parent belt. They predict based on their models that the disk will be bluer beyond 43 AU due to this population of small, barely bound grains, while the inner disk contains a mixture of grain sizes. Strubbe and Chiang (2006) did not discuss whether there would be any expected gradient in the size distribution interior to 43 AU.

Chiang and Fung (2017) modeled the moving features reported by Boccaletti et al. (2015) as being formed by avalanches triggered within the birth ring by the intersection with a secondary ring. The avalanches produce small, ~0.1 micron particles whose dynamics are influenced by the stellar wind.

*Far IR and sub-mm*

Matthews et al. (2015) combined Herschel, JCMT/SCUBA-2, and ALMA measurements to produce a far-IR to mm wavelength spectral energy distribution (SED) of AU Mic. The thermal emission can be fit by a single blackbody at 53 ± 2 K; however, as Matthews et al. (2015) point out, this fit to the spectral information cannot account for the observations of an extended halo containing dust at multiple temperatures. They found that the data can also be fit by a more physically realistic two-temperature-component model with a cold planetesimal belt (39 K) and a warmer halo.

Matthews et al. also found the spectral index, the power law index of the drop-off in flux with wavelength, is shallow in the far-IR and mm wavelengths with a value of around 1.5-2.0. The long-wavelength spectral index is related to the power-law size-distribution index (Draine 2006) with a shallow spectral index implying a shallow grain-size power-law index. Hence, Matthews et al. (2015) note that the spectral index derived from their compiled SED is consistent with the shallow size distribution power-law index of *q* = 3.0 found by Wilner et al.(2012), which is not the value expected of a steady-state collisional cascade. This estimate is in contrast to the grain size distribution implied from the blue spectral index in visible and near IR scattered light, which underestimates the number of large grains and thus the thermal IR emission. From the AU Mic SED fit, they estimate a 1mm dust grain mass of 0.01 $M_{Earth}$ ($5.97 \times 10^{25}$ g). This is similar, but slightly lower than both the previous estimate from Liu et al. (2004) of 0.011 $M_{Earth}$ ($6.57 \times 10^{25}$ g) and the estimate of MacGregor et al. (2013) of $7 \times 10^{25}$ g.

*Scattering model*

Here we present a MCMC model to retrieve grain properties from HST/STIS scattered light data of the AU Mic debris disk presented by Lomax et al. (2018). The light-scattering response is dependent on the shape of the dust grains, their optical properties, and their orientation with respect to the star and the observer. Hence, we need a model that can determine if these spectroscopic changes are due to the sampling of different portions of the light SPF or changes in the size or composition of the dust grains themselves. Thus far, spherical grains have not been able to provide an independently consistent match to both visible/near infrared scattered light and thermal infrared emission data of debris disks (Matthews et al 2015; Rodigas et al. 2017).

## 3. Disk model

Models of scattered light observations of debris disks often assume compact, spherical particles (e.g. Kruegel and Siebenmorgen 1994), although other grain shapes such as ellipsoids and distributed hollow spheres have been considered (e.g. Min, Hovenier, and de Koter 2005). In this work we use the discrete dipole approximation (DDA) method (Draine and Flatau 1994; Zubko, Petrov, Shkuratov, et al. 2005) to calculate scattering efficiencies for realistic grain shapes known as agglomerated debris particles (Zubko, Petrov, Shkuratov, et al. 2005) and use these to model the optically thin disk surrounding AU Mic. We compare the dust properties retrieved using the agglomerated debris grain model with both compact and porous spheres. Because the algorithm used to generate irregular agglomerated debris particle dust grain shapes yields roughly 70-80% porosity with a mean value of 76.4% (Zubko et al. 2015), we compare the results from the agglomerated debris particles to the Lorenz-Mie calculations with 76.4% porosity.

*Fixed stellar and disk parameters*

AU Mic is a spectral type dM1e star located 9.72 pc from earth (Gaia collaboration 2018) with mass $M_* = 0.5 M_\odot$, radius $R_* = 0.93 R_\odot$, effective temperature $T_* = 3300$ K, and luminosity $L_* = 0.1 L_\odot$ (Kalas et al. 2004; Metchev et al. 2005; Arnold et al. 2019).

To limit the number of parameters to be fit by our model we define the disk shape using the inner and outer edge, opening angle, and surface density profile derived from ALMA data (MacGregor et al. 2013). The surface density of the disk increases with distance ($\Sigma \sim r^{2.8}$) up to the ring (MacGregor et al. 2013). and then falls off ($\Sigma \sim r^{-1.5}$) according to Matthews et al. (2014). The disk height as a function of distance is calculated using the opening angle. For a list of fixed disk parameters see Table 1. It is not necessarily the case that the smaller grains share the same spatial distribution as mm and larger grains; however, fixing the spatial distribution of the grains allows us to work with a smaller number of parameters, focusing on composition and size distribution. Moreover, it is difficult to distinguish whether differences in the spatial distribution of grains observed at different wavelengths are real or due to differences in detectability arising from the effects of different instrument sensitivities and grain optical properties (Augereau and H. Beust 2006; Hughes et al. 2018). So, we take this as our initial assumption and revisit it below in the discussion.

*Dust grain shape model*

To calculate the light-scattering properties of irregularly shaped agglomerated dust grains we use an implementation of the DDA method that is described by Zubko et al. (2010) and briefly summarized here. Particles are built up from electric dipoles placed within a 3D cubic lattice of points. The agglomerated debris particle shape (an example of which is shown in Figure 1) is the result of building up randomly shaped solid blocks of material (aka monomers) interspersed with pore spaces of a similar size around randomly placed seed particles. The seed particles are randomly distributed within a circumscribing sphere of size parameter $x = 2\pi a/\lambda$, where $a$ is the radius of the circumscribing sphere and $\lambda$ is the wavelength of the incident light. The average porosity for these particles relative to a solid sphere of the same size parameter $x$ is 76.4% (Zubko et al. 2015). For a more detailed description of the prescription for generating agglomerated debris particles, including how an appropriate number of dipoles for a given grain size is determined and the resulting computational requirements, see Arnold et al (2019).

The scattering properties are averaged over at least 500 randomly generated particles placed in random orientations relative to the direction of incident light. More particles are added as necessary until the addition of new particles does not change the standard deviation of the degree of linear polarization by more than 1% at each scattering angle. This averaging of randomly generated particles ensures that the SPFs are not orientation dependent and are therefore representative of the bulk dust properties. Agglomerated debris particles have significantly different SPFs from spheres or other commonly used approximations such as Henyey-Greenstein functions, particularly in the backscattering region (see Figure 2 for comparison). The most compelling reason for using the agglomerated debris particles in these modeling efforts is they are the only model particles whose scattering properties have been demonstrated to reproduce experimentally measured scattering properties of dust having the same optical properties and size distributions (Zubko et al., 2013; Videen et al., 2018; Zubko, 2015).

*Dust grain composition*

To accomplish the task of understanding whether and how the AU Mic dust grain population varies across the disk, we developed a model capable of fitting disk properties from a multi-wavelength, spatially resolved dataset. We first generated tables of the scattering efficiencies and phase functions for both spheres and agglomerated debris particles (Zubko et al. 2015) as a function of size parameter and complex index of refraction. The sphere scattering properties were calculated using Lorenz-Mie theory, while those of the agglomerated debris particles were calculated using the DDA method described in the previous section. Effective refractive indices for the porous spheres were calculated using the Bruggemen mixing rule (Bruggeman 1935).

To represent debris-disk dust grain compositions, we use the wavelength-dependent optical constants, i.e. the real and imaginary components of the index of refraction ($m = n + i\kappa$) of components common in both debris disk and solar system dust grains. These components include the following: astronomical silicate (Draine and Lee 1984; Draine 2003 b,c) with the chemical formula $MgFeSiO_4$, amorphous carbon (Zubko 1996), and water ice (Henning and Stognienko 1996, online database), tholin (Khare et al. 1984), troilite (Henning, online database), and metallic iron (Henning, online database). We then build a 3D model of the disk using previously measured inclination, inner and outer edge, opening angle and surface mass density. For each voxel (i.e 3D pixel) in the 3D disk model, we calculate the observed phase angle.

*MCMC Model*

We implement the widely used Python MCMC package (Foreman-Mackey et al. 2013), dubbed "emcee," to fit the telescopic spectra from Lomax et al. (2018). We chose the Lomax et al. (2018) data because these are multi-spectral with higher spectral and spatial resolution than previous photometric measurements that averaged over large spatial areas.

To reduce the computational burden, we bin the spectra to a lower wavelength resolution (0.05 microns) and only fit 4 of the 8 available projected distance points (10 AU, 20 AU, 30 AU and 40 AU) from Lomax et al. (2018). For each of these projected separations, the NW and SE sides of the disk were averaged. We use the surface rising density measured by ALMA up to the main dust belt and then the falling surface density of the halo to get the number density of grains for each voxel. This number density is then normalized by the total visible mass parameter being explored by the MCMC as follows, where $N$ is the total number of grains, $mexp$ is the mass exponent, $a$ is the grain radius, P is the porosity, $\rho_i$ is the density of each component, and $f_i$ is the volume fraction of each compositional component.

$$N = 10^{mexp} \Big/ \int_{a_{min}}^{a_{max}} \frac{4}{3}\pi\left(\sum_i P_i f_i \rho_i\right) a^3 n_a \, da$$

The parameters we fit are minimum grain size, grain size distribution power law slope, total disk mass, and the volume fraction of each compositional component (Table 2). Due to computational limitations, the maximum grain size is not a free parameter and set to a radius of 100 microns. This is smaller than the maximum size determined from the mm emission model, but it is not necessary to incorporate such large sizes in the scattering model as these large grains are responsible for a negligible fraction of the scattered light (Zubko et al., 2020). We use a uniform prior for each parameter. The minimum grain size is allowed to range from 0.01 µm to 1 µm, the power law distribution slope from 1 to 5, the mass exponent from 22 to 26 (i.e. the explored mass range is $10^{22}$ to $10^{26}$ g), and the volume fraction of each of the compositional components from 0 to 1. To explore this parameter space, we set the number of walkers to 1000. When calculating the total disk mass, we only fit up to the maximum dust grain size in the scattering model. The total volume fraction is subject to the Dirichlet condition; i.e., they must sum to 1.

For each set of tested parameters, the scattering properties are retrieved from the relevant grain sizes and wavelength-dependent refractive indices. We then use these in combination with the previously calculated phase angles to derive the flux ratio contributed by each voxel. We performed this retrieval for three different grain models: compact spheres, porous spheres, and agglomerated debris particles. Due to the fact that troilite and iron are highly absorbing, they have large values of imaginary refractive index, and we do not have DDA results for ADPs of these values.

. For these components we default to a spherical model and model the rest of the components as agglomerated debris particles. While this is not ideal, the majority of the dust is not made of metallic components, so we will still be able to gain an understanding of how dust shape influences the retrieved composition. The model disk is projected into a 2D image by summing the flux contributions from the 3D model to each 2D pixel. We then extract the modeled spectra and compare these with the Lomax et al. (2018) data. The MCMC then uses these fits to search the parameter space.

## 4. Results and Discussion

The spectra for all four projected separations are fit simultaneously over all wavelengths (Figures 3-5). In doing these simultaneous fits, we assume that the grain size distribution and composition do not have a substantial radial variation. We also did separate fits at each projected distance using the agglomerated debris grain model. Table 2 gives the median and 1σ values for the posterior distribution of all MCMC models. Our model results show that the observed radial change in spectral slope for AU Mic *does not require* a difference in the size distribution or composition of the grains between the inner disk and the main dust belt at 40.3 AU.

*Dust size distribution*

We find that the change in spectral slope as a function of projected distance can be modeled solely with the SPF's of a single size distribution. All three dust grain shape models have a small modeled minimum grain size of roughly 0.2 microns. Notably though, the MCMC posteriors for the Lorenz-Mie particles (see Appendix) show a small, second cluster in the posterior distribution at a larger size 0.8 microns.

The Lomax et al. (2018) STIS data do not extend far enough beyond the 40.3 AU dust belt to evaluate whether the halo is predominately made of small grains. When considered separately, the median fit to the minimum grain size at 40 AU is much smaller than for any of the other projected distances, at 0.08; micron, however, the 1σ of the distribution is large and overlaps with those of the other projected distances and the simultaneously fit dust grain models.

The power-law size-distribution exponents vary between the three grain shape models with the median of the posterior of $q$ = 3.72 (-0.28,+0.43) for the compact spheres, $q$ = 4.57 (-0.40, +0.28) for the porous spheres, and $q$ = 4.20 (-0.36, +0.51) for the agglomerated debris particles. The fits for the porous spheres have little overlap with the other two grain shapes, while there is some overlap in the 1σ for the compact spheres and the agglomerates. Adding porosity appears to favor a power-law exponent tilted towards smaller grain sizes.

No matter which grain model is used, our estimated power-law exponent is greater than the 3.5 expected for a collisional cascade (Dohnanyi 1969) which is consistent with previous fits of AU Mic's surface-brightness profile at visible and near IR wavelengths. For example, Kalas, Liu, and Matthews (2004) fit the power-law indices for the disk midplane at optical wavelengths with 3.6 for the NW and 3.9 for the SE. Augereau and Beust (2006) predict a small minimum grain size of 0.1 um and a power-law index of $q$ = 3.74 (NW) and 4.01 (SE).

This is in contrast to the estimated power-law index for mm *grains* $q$ < 3.31 (MacGregor et al 2016). The power-law exponent can deviate from 3.5, depending on a number of factors not considered by Dohnanyi (1969), such as a size-dependent velocity dispersion and the transition between the gravity and tensile strength regimes (e.g. Pan and Sari 2005; Pan and Schlichting 2012; Gáspár et al. 2012). These more detailed simulations show that while a power law is an overall good approximation to the size/mass spectrum, waves appear both at the boundary between the strength and gravity regimes and at the smallest sizes where grains are affected by radiation forces and stellar wind (Krivov et al. 2006; Wyatt et al. 2011; Pan and Schlichting 2012; Gáspár et al. 2012). While Wyatt (2011) finds that P-R drag causes a turnover in the distribution at smaller sizes, Gáspár et al. (2012) do not find this effect to be so prominent. It is therefore unclear if this can explain the disagreement between the power laws derived from visible /near IR data and mm data.

*Disk mass*

Previous estimates of the total disk mass based on the size distribution retrieved from scattered light do not match estimates based on longer-wavelength data. Augereau and Beust (2006) estimated a total mass for the visible dust of roughly 1-5 $x10^{-4}$ $M_{Earth}$ (Earth masses). Trying to fit the infrared SED using the visible surface density profile resulted in a power law index *of q* = 3.4 and a total disk mass of 7 $x10^{-3}$ $M_{Earth}$. Similarly, Schüppler et al. (2015) were able to model the SED and ALMA 1.3 mm radial surface brightness profile, but this resulted in a discrepancy with observations in the visible, which show a bluer disk and lower degree of polarization inward of 40 AU. They posit that the discrepancy may be due to using a spherical grain model.

All simultaneous fits for the different grain-shape models, as well as the separate projected distance fits using the agglomerated debris particles give a similar estimate for the visible disk mass ranging from $1x10^{24}$ g (1.7 $x10^{-4}$ $M_{Earth}$) to $9x10^{24}$ g (1.5 $x10^{-3}$ $M_{Earth}$) or roughly the mass of Ceres to 3 times the mass of the asteroid belt. Comparing the simultaneous fits for the three dust grain models, solid spheres give a visible dust mass of $6.7x10^{24}$ g, porous spheres give a mass of $1.1x10^{24}$ g, and agglomerated debris particles give a mass of $4.3x10^{24}$ g. This means that when a porous grain model is used, the total overall number of dust grains is increased proportionally in order to fit the data. This mass estimate is only for grains smaller than 100 microns, the maximum size that we used to model the scattered light. To get a sense of whether this matches estimates derived from longer wavelengths, we extrapolate the fit value for the power-law size distribution through the planetesimal size range (10 km). This results in an estimated total mass of $1.4x10^{27}$ g (0.2 $M_{Earth}$) for the compact spheres and $6x10^{24}$ g (0.001 $M_{Earth}$) for the agglomerated debris particles, which is lower than the ALMA derived estimate of $7x10^{25}$ g.

The underestimate from the agglomerated debris particles is similar to the number derived by Augereau and Beust (2006) and in both cases is due to extrapolating a size distribution more heavily weighted towards very small grains. The overestimate in the compact-sphere case is likely due to the fact that the total flux is also overestimated by that model.

*Dust composition*

While the dust size distribution and mass are well constrained, there is substantial degeneracy in the composition, even simultaneously fitting over multiple wavelengths and projected distances. The estimated silicate volume fraction is similar between the compact spheres and the agglomerated debris particles 28% (+24%, -19%) vs 32% (+22%, -21%), respectively, while it is not well constrained for the porous spheres 13% (+16%, -10%). When looking at the posterior distributions (Appendix), the amount of astronomical silicate is correlated with water ice in the solid sphere model and tholins in the agglomerate model.

The volume fractions of amorphous carbon and tholins are correlated in the solid-sphere model and all three of amorphous carbon, tholins and water ice are correlated in the agglomerated debris model, making it difficult to place an estimate for these three components. Similarly, the two metallic phases, troilite and iron, despite having different spectral slopes, are highly correlated with each other as well as with amorphous carbon. This is unsurprising given that these phases are all highly absorbing.

Given these results, it is not possible to give exact quantities for the various components, but the fits do provide useful constraints. It is likely that the disk contains a substantial fraction of silicates of around 30% by volume. None of the models favor a large total abundance of highly absorbing phases (amorphous carbon, troilite, and iron). It is unlikely that these phases comprise more than 20-30% of the dust composition. Given that the proportion of these components is similar for both the compact spheres and the two types of porous grains, it is clear that there is some information about composition that can be extracted separate from grain morphology. Moreover, the blue color of the disk in the visible and near IR is a result of the small grains residing in the disk and does not require the inclusion of water ice.

*Polarization*

The degree of linear polarization provides an additional constraint on modeling. Graham et al. (2006) made measurements of the polarization across the disk of AU Mic. We used our median MCMC fit to model the degree of linear polarization as a function of radial distance (Figure 6). While the more realistic agglomerated debris particles produce a linear polarization closer to the measured polarization, it is not a better fit than the porous, icy grain model of Graham et al. (2006). The polarizations resulting from using Mie theory are significantly different than a that of the other modeling results and the experimental data.

## 6. Summary

We modeled the AU Mic debris-disk scattered light at wavelengths of $\lambda$=0.5-1 microns to better understand the range of dust properties that are able to fit the data. The MCMC model allowed us to test not only whether the disk can be fit with a single set of dust size distributions and compositions, but also allowed us to explore a broader range of compositions than previous works.

We have found that a small minimum grain size (0.2 um) is necessary to fit the blue slope observed by Lomax et al. (2018) in the inner disk. This is consistent with the fact that radiation pressure is not expected to be effective at removing dust from the system, and the minimum grain size is instead set by the stellar wind. We have also found that fitting the wavelength-dependent flux ratio at 40 AU requires adding a halo with a decreasing surface density that extends past the ALMA emission detection, consistent with the results of Matthews et al. (2015).

We also compared the retrieved dust properties for three different models: compact spheres, porous spheres, and agglomerated debris particles. Both spherical grains and the agglomerated debris particles give a similar minimum grain size within the disk of 0.17 micron for both the compact and porous spheres and 0.23 microns for the agglomerated debris particles, respectively. However, the spherical models have a secondary cluster of solutions at 0.8-1.0 microns.

The grain size distribution power law index also varies between the dust grain models, with a value *of* $q$ = 3.7 for the compact spheres, $q$ = 4.2 for the agglomerated debris particles, and $q$ = 4.6 for the porous spheres. These agree with previous estimates based on visible/near infrared data and are overall steeper than is measured for the millimeter.

We have found that the observed radial change in disk color of the AU Mic debris disk can be explained by the fact that different portions of the SPF are being observed at different projected distances from the star. For the observed separations of 10-40 AU, there is no need to invoke a change in either the dust grain composition or size distribution; however, to match the overall flux ratio at 40 AU, it is necessary to have a halo of material that extends past the radius where the disk is observed by ALMA.

There is a great deal of degeneracy in the modeled dust composition. Compact spheres and agglomerated debris particles give a similar silicate fraction of 0.28 and 0.32, respectively, while MCMC was not able to constrain the silicate content for the porous sphere model. Water ice and tholin content are not well constrained for any dust grain model. The fractions of amorphous carbon, troilite, and metallic iron are all correlated with a strong preference for a total opaque fraction of less than 0.3, except in the porous-sphere case where the total of these three phases is higher. While it may not be possible to determine the exact breakdown, this does provide an upper limit for opaque components. Overall, it is possible to fit the STIS spectra with a wide variety of compositions.

**Acknowledgments**
Support for Program number HST-AR-14590.002-A was provided by NASA through a grant from the Space Telescope Science Institute, which is operated by the Association of Universities for Research in Astronomy, Incorporated, under NASA contract NAS5-26555. This project is supported by NASA ROSES XRP, grant NNX17AB91G.

**Tables & Figures**

Table 1: Fixed disk properties in MCMC model

| **Description** | **Parameter** | **Value** |
|---|---|---|
| Disk inner edge | $R_{inner}$ | 8.8 AU |
| Disk outer edge | $R_{outer}$ | 40.3 AU |
| Halo cutoff | $R_{halo}$ | 80 AU |
| Surface density exponent disk | $x_{disk}$ | 2.82 |
| Surface density exponent halo | $x_{halo}$ | -1.5 |
| Opening angle | 2δ | 2.3° |
| Inclination angle | $i$ | 89.5° |

Table 2: Fit of disk dust size distribution and composition.

| **Description** | **Parameter** | **Range explored** | **Median (-/+1 σ)** | | |
|---|---|---|---|---|---|
| | | | Compact Sphere | Porous sphere | Agglomerate |
| **Minimum grain radius (micron)** | $a_{min}$ | 0.01-1.0 | 0.17 (+0.06, -0.02) | 0.17 (+0.01, -0.01) | 0.23 (+0.05, -0.06) |
| **Size distribution power law exponent** | $a_{exp}$ | 1.0 to 5.0 | 3.72 (-0.28, +0.43) | 4.57 (-0.40, +0.28) | 4.20 (-0.36, +0.51) |
| **Mass exponent (g)** | $m_{exp}$ | 15.0-26.0 | 24.83 (+0.53,-0.47) | 24.03 (+0.23, -0.10) | 24.63 (+0.31, -0.20) |
| **Silicate volume fraction** | $f_{si}$ | 0-1.0 | 0.28 (+0.24, -0.19) | 0.13 (+0.16, -0.10) | 0.32 (+0.22, -0.21) |
| **Amorphous carbon volume fraction** | $f_{ac}$ | 0-1.0 | 0.09 (+0.12, -0.07) | 0.15 (+0.18, -0.11) | 0.08 (+0.11, -0.06) |
| **Water ice volume fraction** | $f_{wi}$ | 0-1.0 | 0.23 (+0.19, -0.14) | 0.12 (+0.17, -0.09) | 0.15 (+0.20, -0.11) |
| **Tholin volume fraction** | $f_{th}$ | 0-1.0 | 0.11 (+0.16, -0.08) | 0.12 (+0.16, -0.09) | 0.21 (+0.23, -0.15) |
| **Troilite volume fraction** | $f_{tr}$ | 0-1.0 | 0.08 (+0.11, -0.06) | 0.15 (0.14+, -0.10) | 0.07 (+0.07, -0.05) |
| **Metallic iron volume fraction** | $f_{fe}$ | 0-1.0 | 0.07 (+0.09, -0.05) | 0.15 (0.14+, -0.10) | 0.04 (+0.05, -0.03) |

Table 2 cont'd: Fit of disk dust size distribution and composition.
**Median (-/+1 σ)**

| Agglomerates 10 AU only | Agglomerates 20 AU only | Agglomerates 30 AU only | Agglomerates 40 AU only |
|---|---|---|---|
| 0.20 (+0.11, -0.15) | 0.39 (+0.10, -0.15) | 0.19 (+0.06, -0.13) | 0.07 (+0.16, -0.04) |
| -4.36 (+0.49, -0.48) | -4.22 (+0.60, -0.56) | -4.23 (+0.74, -0.57) | -3.90 (+0.45, -0.65) |
| 24.67 (+0.41, -0.25) | 24.97 (+0.62, -0.33) | 24.65 (+0.89, -0.34) | 24.86 (+0.72, -0.53) |
| 0.17 (+0.22, -0.12) | 0.14 (+0.20, -0.10) | 0.25 (+0.25, -0.18) | 0.15 (+0.24, -0.11) |
| 0.09 (+0.14, -0.07) | 0.09 (+0.14, -0.07) | 0.07 (+0.11, -0.05) | 0.11 (+0.17, -0.08) |
| 0.15 (+0.23, -0.11) | 0.14 (+0.21, -0.11) | 0.18 (+0.25, -0.13) | 0.14 (+0.23, -0.11) |
| 0.14 (+0.21, -0.10) | 0.18 (+0.22, -0.13) | 0.16 (+0.23, -0.12) | 0.13 (+0.21, -0.10) |
| 0.10 (+0.16, -0.08) | 0.10 (+0.16, -0.08) | 0.08 (+0.10, -0.06) | 0.10 (+0.17,-0.07) |
| 0.10 (+0.15, -0.07) | 0.11 (+0.17, -0.08) | 0.06 (+0.08, -0.04) | 0.08 (+0.18,-0.06) |

Figure 1: Examples of agglomerated debris particles.

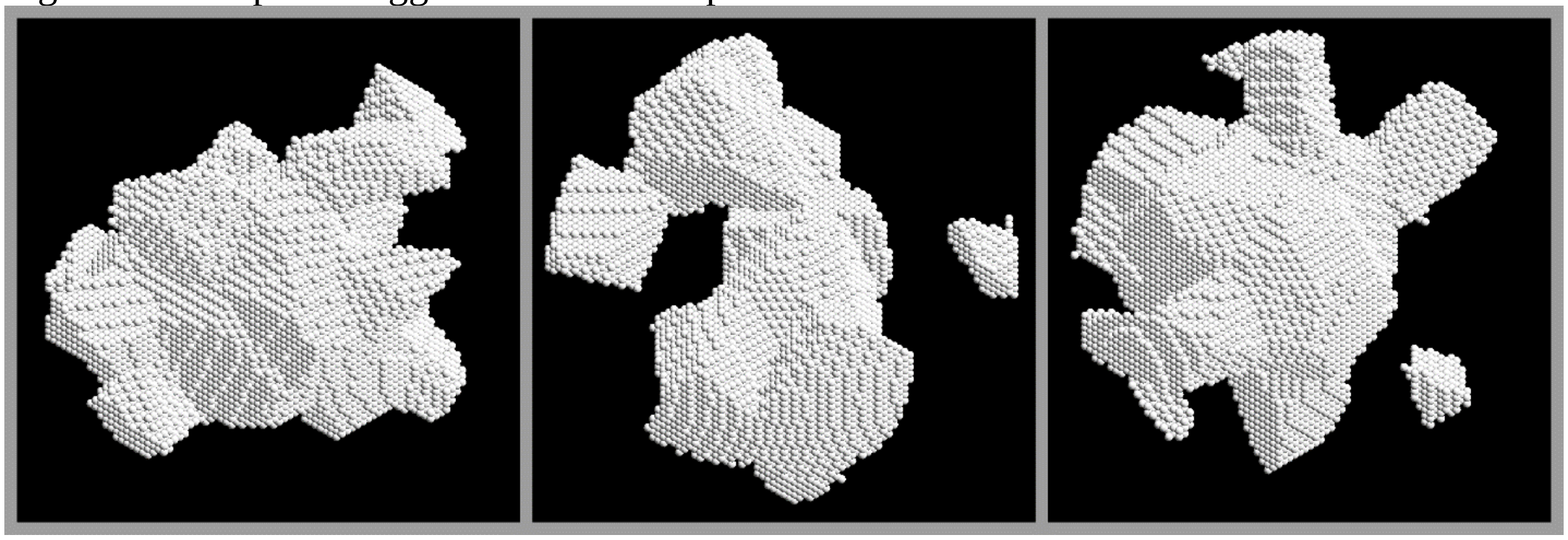

Figure 2: Comparison of SPFs of spheres (solid bold lines), agglomerated debris particles (dashed lines), and Henyey-Greenstein functions (light grey lines) for two different power-law size distributions. The complex index of refraction chosen represents values typical of silicates at visible wavelengths. The particle size parameters range *from x* = 0.1 to 48, which correspond to spheres of diameter 32 nm to 15.23 microns for an incident beam of 1 micron.

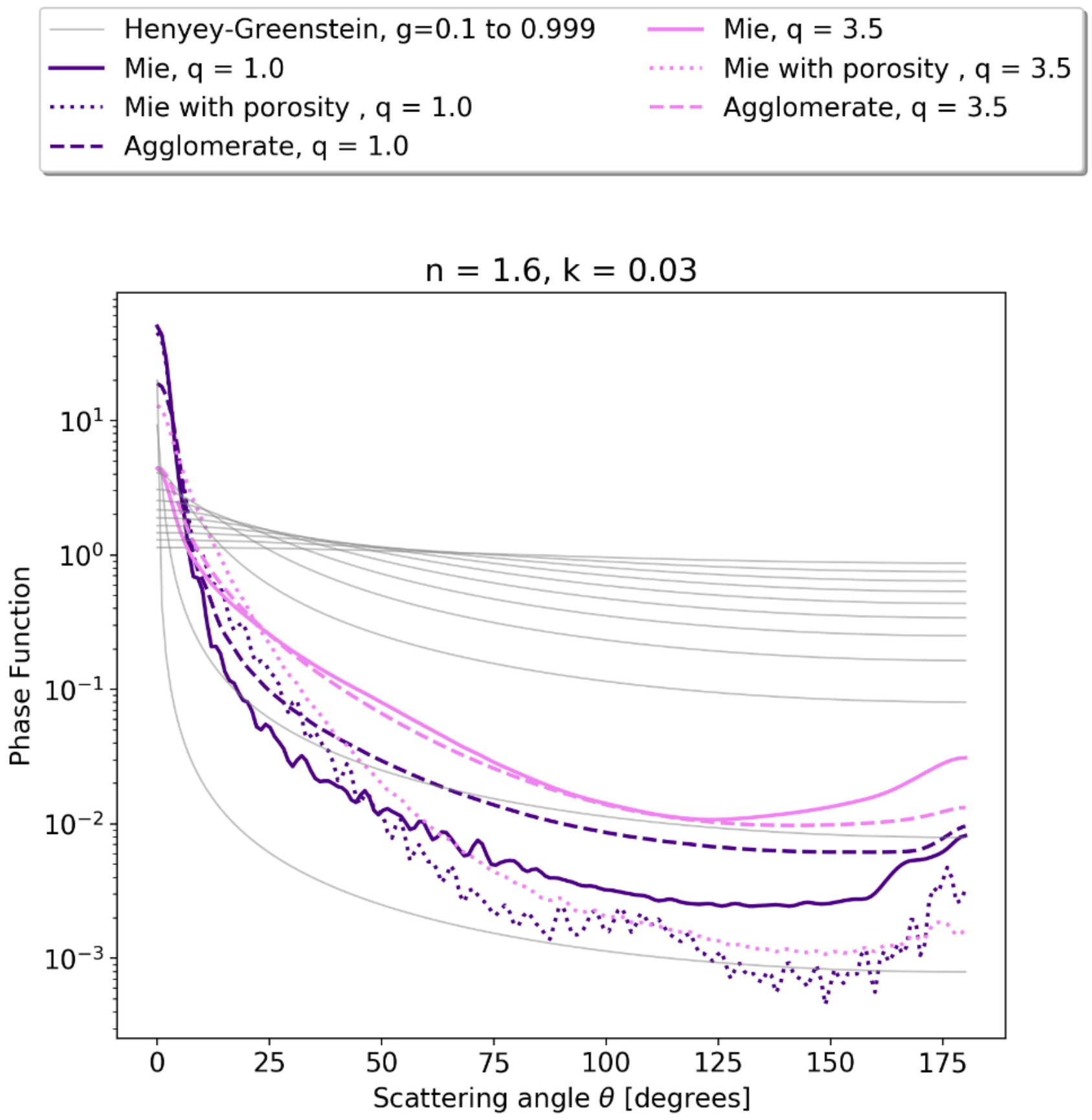

Figure 3: Fit to reduced spectral resolution STIS data for compact spheres.

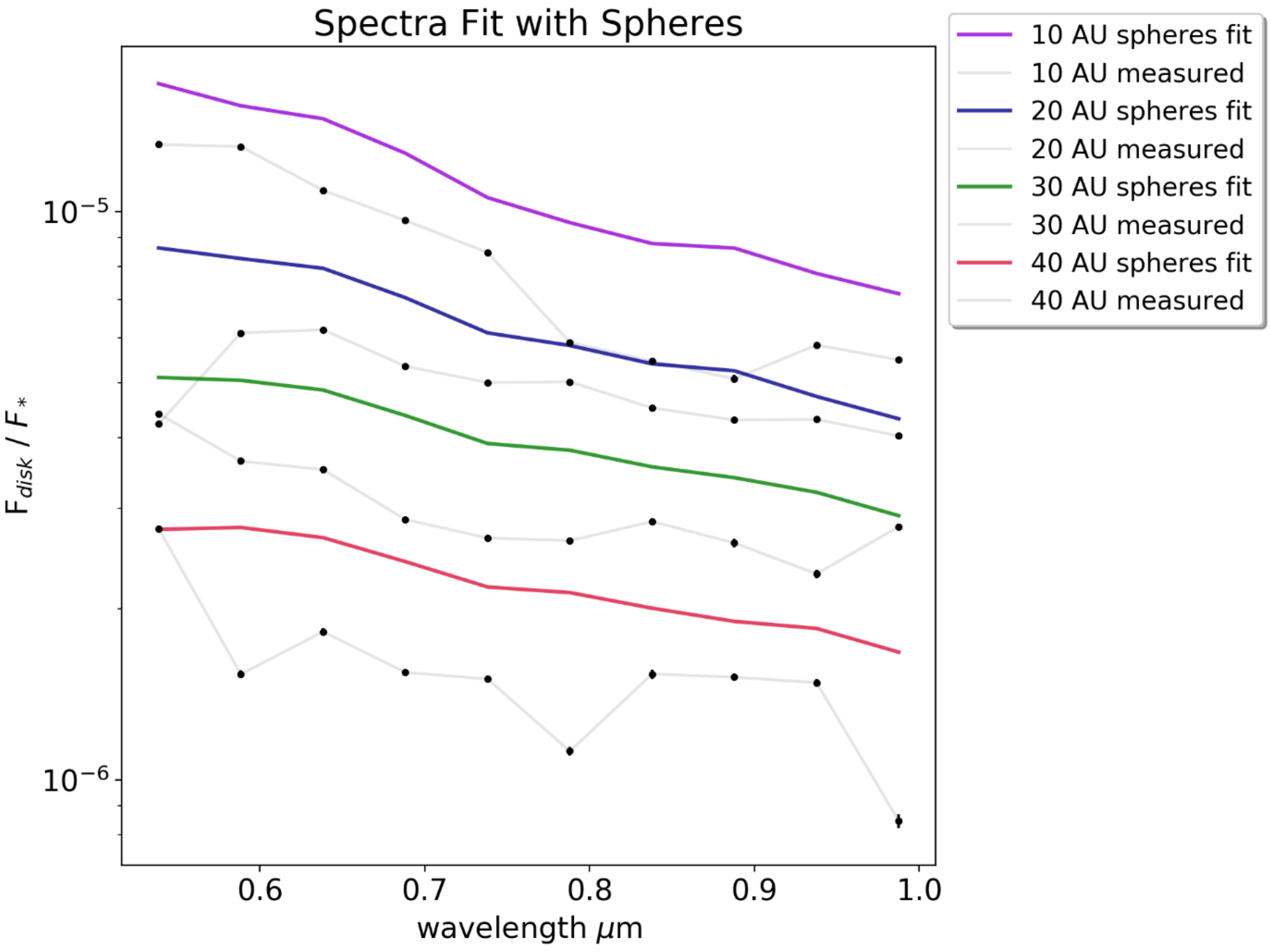

Figure 4: Fit to reduced spectral resolution STIS data for porous spheres.

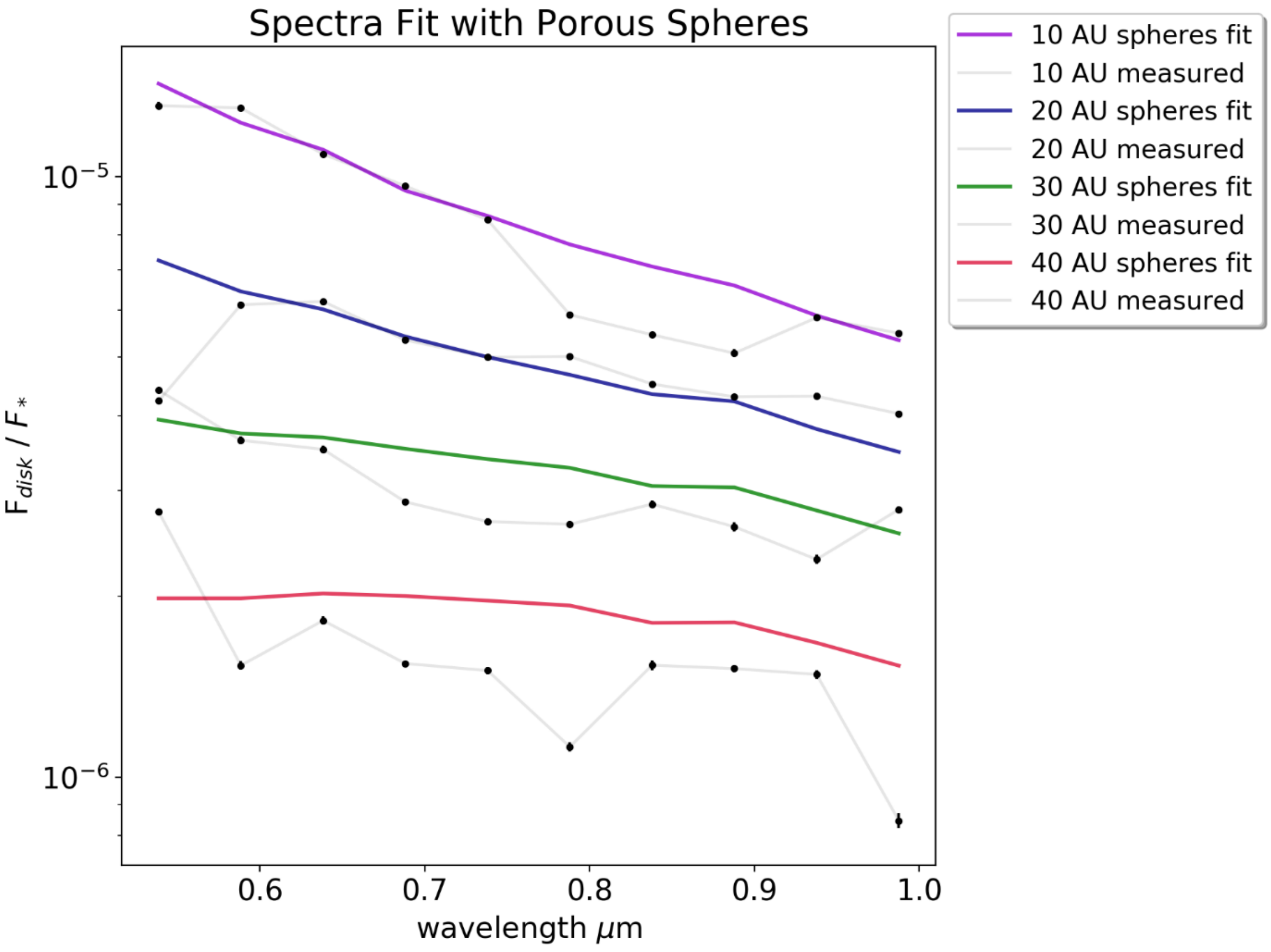

Figure 5: Fit to reduced spectral resolution STIS data for agglomerates

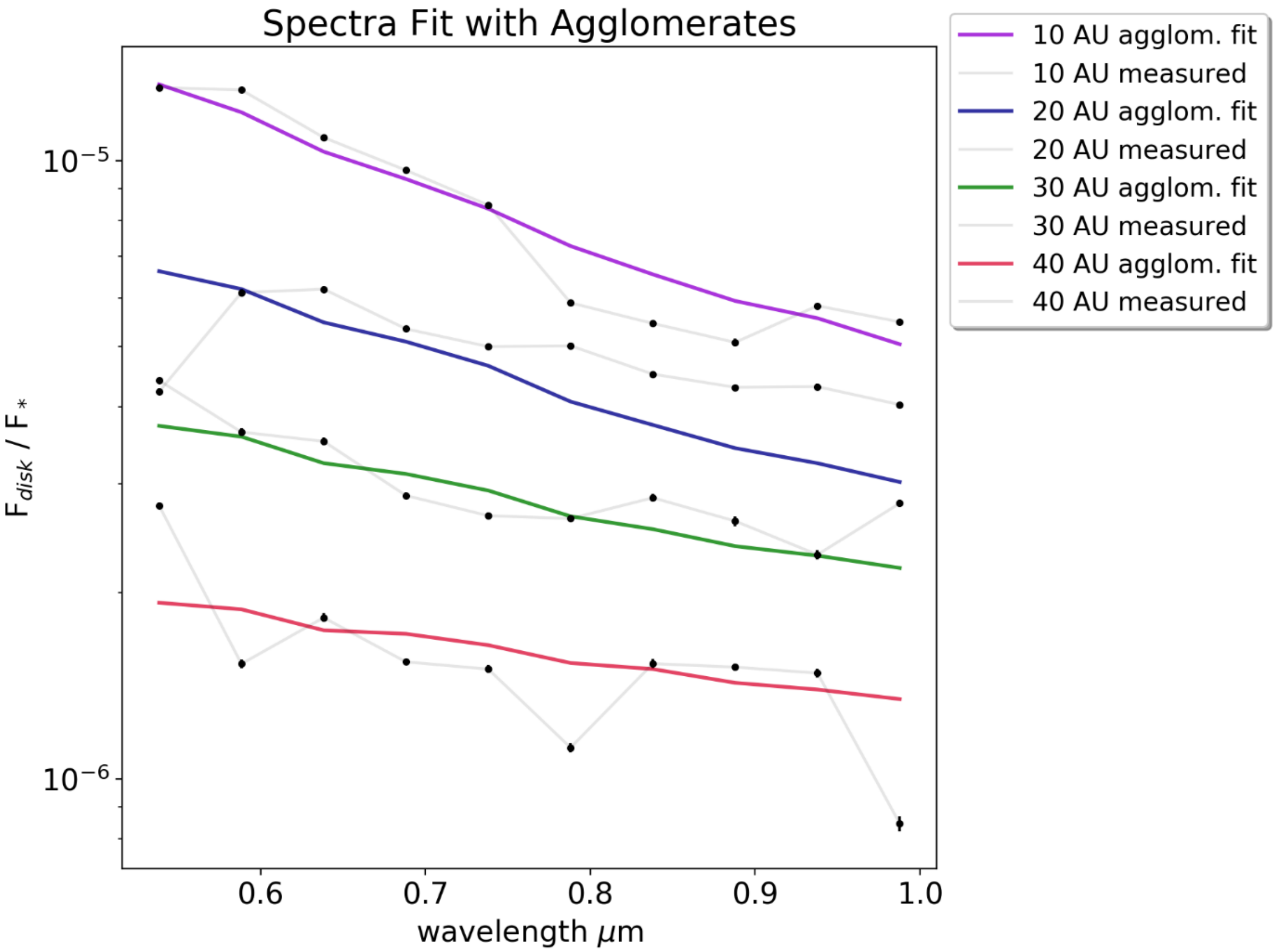

Figure 6: Degree if linear polarization as a function of projected separation modeled using our median MCMC fits for each grain model. We compare this to the Graham 2007 data (pink dots).

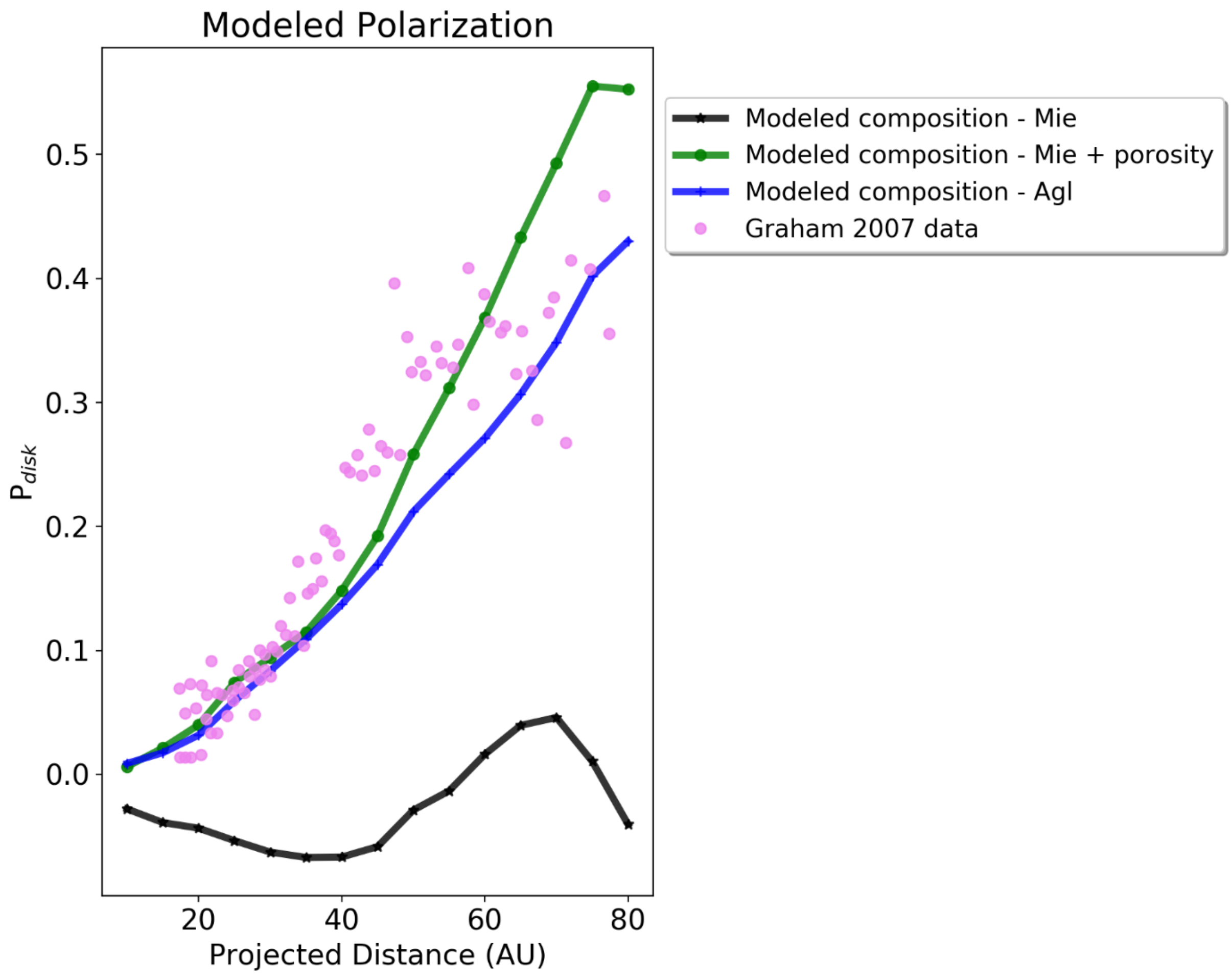

Figures I'd like to add
- FIR SED plots
- SPF comparison for each model

**Appendix 1:** MCMC corner plots

## A1.1. Compact spheres

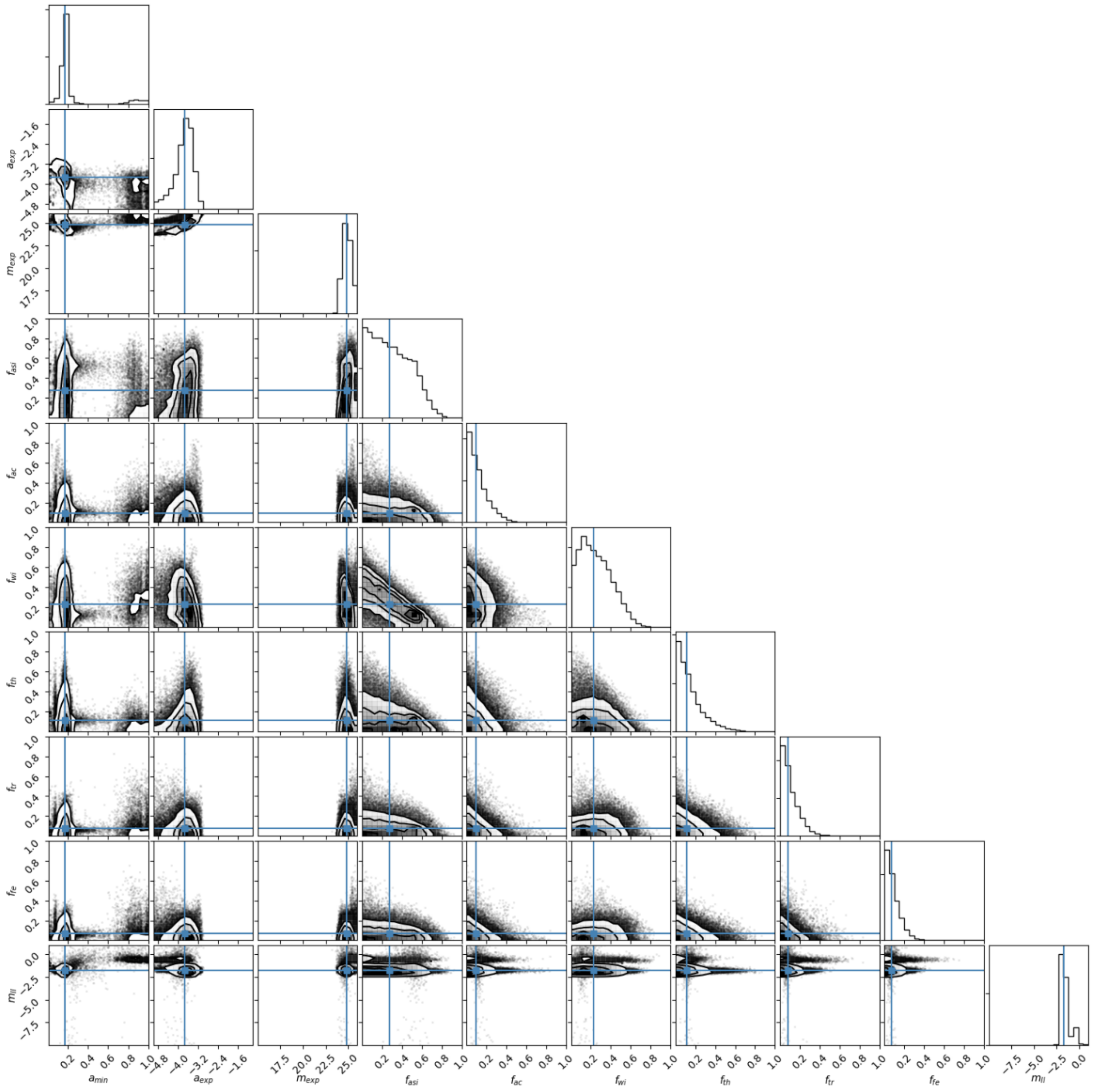

## A1.2. Porous Spheres

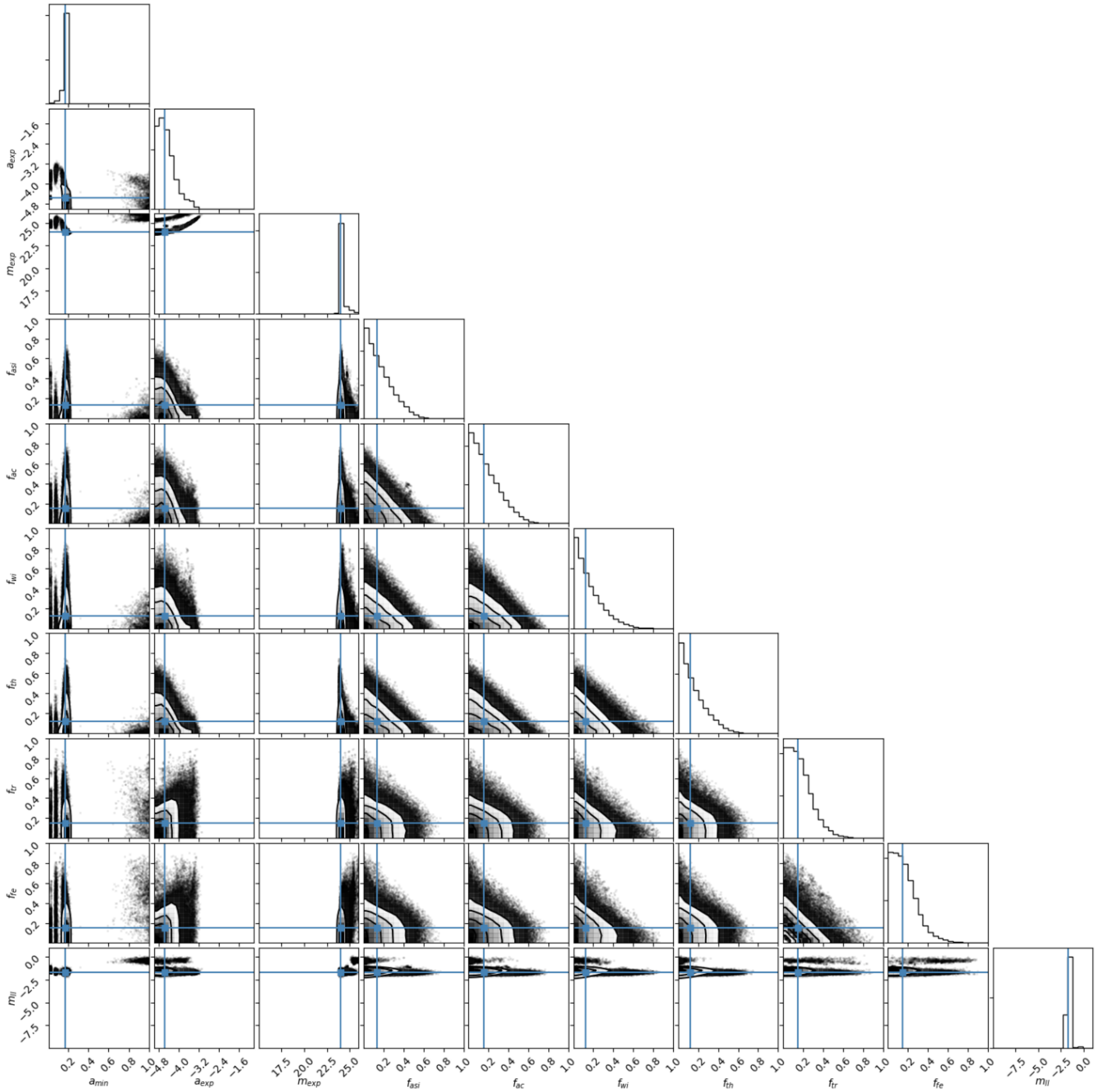

A1.3. Agglomerates

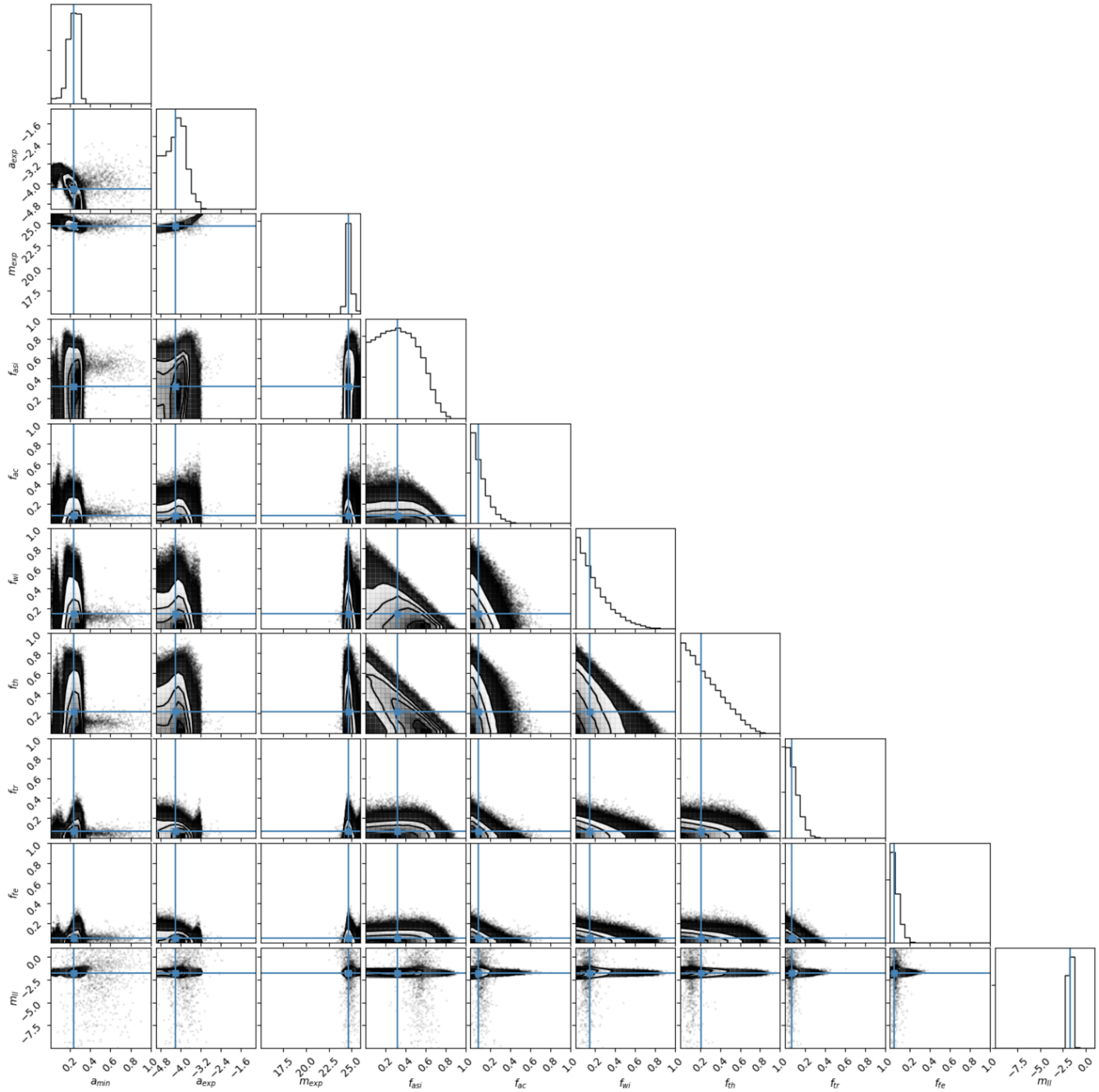

**Appendix 2:** Description of the tables of scattering phase functions and efficiencies.

To make the best use of the DDA scattering matrix calculations of the agglomerated debris particles, we interpolated over the available refractive indices and size parameters and created lookup tables from these interpolated values. The refractive indices were interpolated using the methods given in Zubko et al. (2018). After interpolating the over refractive index for each size parameter, we then applied a cubic interpolation over size parameter for the scattering efficiency. For the phase function, a cubic interpolation for small scattering angles (<15 degrees) and a linear interpolation for large scattering angles worked best. These interpolations were used to extrapolate to larger values of size parameter. The angular resolution of the phase functions was set to 3 degrees to speed up the MCMC code. We then made a similar lookup table for the Mie spheres over the same range of refractive index, size parameters, and scattering angles. Unlike the agglomerated debris particles, Mie scattering calculations are not time consuming, so these were calculated directly with no interpolation.

The tables of scattering phase functions and efficiencies that are given in the supplemental material are in the Hierarchical Data Format (.hdf5). These table contain arrays with size parameter, complex refractive index, scattering efficiency, and phase function.